# Localisation of a wave-function by superposition of different histories

D.J. Dunstan
Physics Department, Queen Mary, University of London,
London, E1 4NS.

**Abstract:** Quantum state diffusion shows how stochastic interaction with the environment may cause localisation of the wave-function, and thereby demonstrates that quantum mechanics need not invoke a separate axiom of measurement to explain the emergence of the classical world. It has not been clear whether quantum state diffusion requires some new physics. We set up an explicit numerical calculation of the evolution of the wave-function of a two-state system under interaction using only the physics explicitly contained in quantum mechanics without an axiom of measurement. The wave-function does indeed localise, as proposed by quantum state diffusion, on eigenstates of the perturbation. The mechanism appears to be the superposition of histories evolving under different Hamiltonians.

# 1. Introduction

The boundary between the quantum world and the classical world is widely acknowledged to occur somewhere on the scale between atoms and mesoscopic objects, but the paradoxes of Schrödinger's Cat (Schrödinger, 1935) and Wigner's Friend (Wigner, 1967) may still be invoked to show that this is arguable. In axiomatic quantum mechanics, the localisation of the wave-function upon measurement is given as an axiom in its own right, for it is not predicted by the unitary evolution of the solutions of the Schrödinger equation (Schrödinger, 1935). Many models of the collapse have been proposed, but none have yet been shown to be the correct explanation. Of these, the quantum state diffusion theory (QSD) and the decoherence programme (DC) provide two of the most attractive, as they rely on what might be called standard physics rather than invoking fantastic concepts such as many worlds, or a role played by consciousness. In both of these models, the collapse is due to stochastic perturbation of the wave-function by interaction with the environment. See the books by Percival (1998) and Guilini *et al.* (1996) for accounts of QSD and of DC, respectively.

Some authors believe that the decoherence programme has solved the problem, and suggest that the textbooks should already be rewritten to reflect this (Tegmark and Wheeler, 2001). Recent experiments by Zeilinger and his group on the quantum interference of buckyballs ($C_{60}$ and $C_{70}$) have pushed the boundary much further up the scale of size than might have been expected (Aendt *et al*, 1999). Indeed, their recent experiments in which they destroy the quantum interference of buckyballs by heating them to 2000K (Hackermüller 2004) might be considered to constitute an experimental demonstration of the role of stochastic interaction with the environment in the collapse of the wave-function.

Both QSD and the DC use advanced mathematical techniques such as master equations with Lindblad operators, to show how the density matrix evolves. This results in a lack of transparency. Thus it is argued that the DC fails to solve the problem of measurement precisely because it shows only how the interference terms vanish (the ensemble density matrix becoming diagonal) and not how the wavefunction localises onto one or the other diagonal elements. All the decoherence programme explains is why we don't see interference (Pearle, 1997; Joos and Zeh, 1985).

It is certainly true that stochastic perturbation will lead to the diffusion of a state-vector in Hilbert space. The peculiarity of quantum state diffusion (QSD) is that the state-vector diffuses towards eigenstates of the perturbation and sticks there, as if to fly-paper (Percival, 1998). This is not how diffusion usually operates – diffusion tends to spread things out rather than concentrate them. In QSD, it is a consequence of the peculiar nature of the Lindblad operators, which are determined by the system-environment interaction Hamiltonian so that the attractors are the eigenstates of the perturbation. It is provable that unitary evolution of the Schrödinger equation is unable to yield such a concentration of the state vector in Hilbert space (Percival, 1998). One may question, therefore, how Lindblad operators achieve localisation – whether they must add some new physics to the axioms of quantum mechanics, or whether localisation is an unexpected consequence of what is already known to occur.

The Second Law of Thermodynamics provides a useful analogy. It is clear that this cannot be derived from classical mechanics, for that has time-reversal symmetry while thermal equilibration does not. Indeed, classical statistical mechanics since Boltzmann has been bedevilled by attempts to derive this asymmetry rigorously through such additional devices as the ergodic hypothesis. Yet we may set up a molecular dynamics simulation of a gas, with only classical mechanics operating. Starting from any non-equilibrium state, it is clear that it will equilibrate whether time is run backwards or forwards (and this was already clear to Boltzmann). It is clear that no new physics is required to account for the thermal equilibration of a system. The calculation we report here is motivated by this analogy.

We have tested the evolution of a quantum system by a process inspired by molecular dynamics simulation. We set up a system whose ground state is an eigenstate of energy, yet which in the classical world localises into an eigenstate of position. We let it interact with an environment, following, as far as we are able, the standard rules of quantum mechanics. We deliberately keep the system as simple as may be, so that analytic solutions are easily obtained. This also enables the effects of quantum entanglement between the system and the environment to be handled analytically. We expected that the calculation would *not* show localisation, and would thereby force the issue of what new physics is needed. To our surprise, however, the wave-function does indeed localise on an eigenstate determined by the nature of the perturbation. It appears that this is not due to stochastic interaction with the environment but is due to the superposition of histories evolving according to different Hamiltonians.

## 2. Quantum Molecular Dynamics Model

*2.1. The Model*

Consider a particle in a pair of potential wells. We may write the two time-dependent wavefunctions for the left-hand and right-hand wells,

$$\Psi_L = \psi_L(\mathbf{r})e^{i\omega_0 t}$$
$$\Psi_R = \psi_R(\mathbf{r})e^{i\omega_0 t} \quad (1)$$

and in the energy basis the (degenerate) ground and first excited states of the system are

$$\Psi_0 = \frac{1}{\sqrt{2}}\Psi_L + \frac{1}{\sqrt{2}}\Psi_R$$
$$\Psi_1 = \frac{1}{\sqrt{2}}\Psi_L - \frac{1}{\sqrt{2}}\Psi_R \quad (2)$$

Introducing a weak coupling lifts the degeneracy and gives a frequency splitting of $\omega_1$ between $\Psi_0$ and $\Psi_1$. The ground and first excited states of the system are then

$$\Psi_0 = \frac{1}{\sqrt{2}}(\Psi_L + \Psi_R) = \frac{1}{\sqrt{2}}(\psi_L + \psi_R)e^{i(\omega_0 - \frac{1}{2}\omega_1)t}$$
$$\Psi_1 = \frac{1}{\sqrt{2}}(\Psi_L - \Psi_R) = \frac{1}{\sqrt{2}}(\psi_L - \psi_R)e^{i(\omega_0 + \frac{1}{2}\omega_1)t} \quad (3)$$

The general state of the system is a superposition, with

$$\Psi = a\Psi_0 + b\Psi_1$$
$$|a|^2 + |b|^2 = 1 \quad (4)$$

and the density matrix in the energy basis $\{|0\rangle, |1\rangle\}$ is

$$\rho_E = \begin{pmatrix} aa^* & ab^* e^{-i\omega_1 t} \\ a^* b e^{-i\omega_1 t} & bb^* \end{pmatrix} \quad (5)$$

Expanding the wavefunction in the spatial basis set $\Psi_L$ and $\Psi_R$, we have time-varying coefficients,

$$\Psi = \alpha(t)\Psi_L + \beta(t)\Psi_R$$
$$= \frac{1}{\sqrt{2}}\left[\left(-ae^{-\frac{1}{2}i\omega_1 t} + be^{\frac{1}{2}i\omega_1 t}\right)\psi_L + \left(ae^{-\frac{1}{2}i\omega_1 t} + be^{\frac{1}{2}i\omega_1 t}\right)\psi_R\right]e^{i\omega_0 t} \quad (6)$$

and the density matrix in the spatial basis $\{|L\rangle, |R\rangle\}$ is, assuming $a$ and $b$ are real,

$$\rho_S = \begin{pmatrix} \frac{1}{2} - ab\cos\omega_1 t & \frac{1}{2}(b^2 - a^2) - iab\sin\omega_1 t \\ \frac{1}{2}(b^2 - a^2) + iab\sin\omega_1 t & \frac{1}{2} + ab\cos\omega_1 t \end{pmatrix} \quad (7)$$

so that the amplitude of the wave-function beats, or oscillates between the two wells. This is shown in Figure 1, curve a. The diagonal matrix elements $|\alpha(t)|^2 = \alpha^*\alpha$ and $|\beta(t)|^2 = \beta^*\beta$ oscillate at the frequency $\omega_1$ and with a beat amplitude that depends on the initial values of $a$ and $b$ (e.g. from 0 to 1 for for $a = b = 1/\sqrt{2}$). Eqn.6 describes well a molecule with two spatial configurations of the same energy, such as ammonia, for which $\omega_1$ is about 24 GHz. For larger molecules or heavier atoms, $\omega_1$ takes smaller values, and in the classical world, very secure localisation occurs in one spatial configuration or the other (Hund, 1927; Woolley, 1976).

We introduce a position-dependent interaction with the environment which raises the energy of the left-hand well or the right-hand well. According to QSD, this is what we require to obtain spatial localisation in one well or the other. The energy eigenvalues and eigenstates for the double-well system with the left-hand well perturbed by $\omega_P$ are given by the diagonalisation of the Hamiltonian,

$$\mathbf{H} = \begin{pmatrix} \omega_0 + \omega_P & \frac{1}{2}\omega_1 \\ \frac{1}{2}\omega_1 & \omega_0 \end{pmatrix} \quad (8)$$

When $\omega_P$ is non-zero, during the interaction with the environment, letting the perturbed well be well $L$, and using a prime to indicate wavefunctions and quantities during the perturbation, Equation 3 becomes

$$\Psi'_0 = u'_L \Psi'_L + u'_R \Psi'_R$$
$$\Psi'_1 = v'_L \Psi'_L + v'_R \Psi'_R \quad (9)$$
$$\Psi' = a'(t)\Psi'_0 + b'(t)\Psi'_1 = \alpha'(t)\Psi_L + \beta'\Psi_R$$

where $\mathbf{u}'$ and $\mathbf{v}'$ are the normalised eigenvectors of the matrix $\mathbf{H}$ of Eqn.8. If the interaction begins at time $t = t_0$, the coefficients $a'(t)$ and $b'(t)$ are obtained by setting $\Psi'(t_0) = \Psi(t_0)$, or $\alpha'(t_0) = \alpha(t_0)$ and $\beta'(t_0) = \beta(t_0)$. The expressions are straightforward to calculate but are somewhat lengthy, so we give the results only for the limiting case of $\omega_1 = 0$ and for a perturbation beginning at $t = 0$. However, the full expressions are used in the numerical calculations which follow. With $\omega_1 = 0$ and $t_0 = 0$,

$$\Psi'_0 = \Psi'_R = e^{i\omega_0 t}\psi_R$$
$$\Psi'_1 = \Psi'_L = e^{i(\omega_0 + \omega_p)t}\psi_L \quad (9)$$
$$\Psi' = \frac{1}{\sqrt{2}}(a+b)\Psi'_0 + \frac{1}{\sqrt{2}}(a-b)\Psi'_1$$

The beat frequency is increased and the beat amplitude is reduced for the duration of the perturbation. At a time $t = t_0 + t_1$ the interaction ceases and the frequency of the perturbed well returns to $\omega_0$ while the beat frequency returns to $\omega_1$. New coefficients $a(t)$ and $b(t)$ are calculated by setting $\Psi(t_0 + t_1) = \Psi'(t_0 + t_1)$. With $\omega_1 = 0$ and $t_0 = 0$,

$$\Psi_{after} = \frac{1}{\sqrt{2}}(a-b)e^{i\omega_p t_1}e^{i\omega_0 t}\psi_L + \frac{1}{\sqrt{2}}(a+b)e^{i\omega_0 t}\psi_R \qquad (10)$$

The perturbation has simply given a phase-shift to the perturbed well – here, the left-hand well – and the outcome of this is only to change the phase and amplitude of the Rabi oscillations of $\alpha^*\alpha$ and $\beta^*\beta$.

*2.2. Random interactions with the environment*

The model of Section 2.1 can be used to simulate the interactions of a single ammonia-like molecule with an environment such as an ideal gas. Collisions with the ideal gas atoms correspond to a succession of such collisions, randomly incident on the left and on the right and with random time intervals between them. Not surprisingly, weakly perturbing or infrequent collisions allow the Rabi oscillations to persist. Stronger and more frequent collisions wipe out the sinusoidal beating of amplitudes between the two wells, and the state vector undertakes a random walk. The spatial density matrix generally has non-zero values both on and off the diagonal, but the ensemble-averaged or time-averaged density matrix becomes

$$\langle \rho_S \rangle = \begin{pmatrix} \tfrac{1}{2} & 0 \\ 0 & \tfrac{1}{2} \end{pmatrix} \qquad (11)$$

It is worth noting that both QSD and the decoherence programme might be thought to predict that the wavefunction should localise under this stochastic interaction with the environment, but there is no true localisation in Eqn. 11 (Pearle, 1997). In fact, we expect no localisation here, as the Schrödinger equation undergoes unitary evolution. The quantum dispersion entropy of the system is not reduced, and so the system must continue to visit all of phase space rather than becoming localised in some part of it. There is nothing new in this part of the calculation, but it is worth noting explicitly that a diagonal density matrix as in Eqn.11 does not imply localisation.

It turns out that we can obtain localisation in this model by introducing a modification to the interaction. The modification is to allow the duration of the interaction to depend on the state of the molecule. A molecule is in a superposition state of the particle being in well $L$ and in well $R$. It is physically reasonable that the evolution of the impact should also proceed in two superposed histories. The duration of the impact will be given two values. In semi-classical language, what we envisage here is that if the incident ideal gas atom approaches from the left, and the particle is in the left hand well, the left-hand well is increased in energy by $\omega_P$ for the time $t_1$ as above. However, if the particle is in the right-hand well, the incident atom travels further before being repelled, and it raises the energy of the (unoccupied) left-hand well for longer, for a time $t_2 > t_1$. There are two histories for the interaction. We therefore match the unperturbed wavefunction $\Psi_{after}$ to the perturbed wavefunction $\Psi'$ according to Eqn.10 at the two times, $t_0 + t_1$ and $t_0 + t_2$, giving $\Psi_{after\,t_1}$ and $\Psi_{after\,t_2}$. The full wavefunction is the superposition of $\Psi_{after\,t_1}$ and $\Psi_{after\,t_2}$, with the appropriate amplitudes, and for $\omega_1 = 0$ and $t_0 = 0$, this gives,

$$\Psi_{after} = \alpha(t_0)\Psi_{after\ t_1} + \beta(t_0)\Psi_{after\ t_2}$$
$$= e^{i\omega_0 t}\left(\tfrac{1}{2}\left((a+b)^2 e^{i\omega_p t_1} + (a^2 - b^2)e^{i\omega_p t_2}\right)\psi_L + a(a-b)\psi_R\right) \quad (13)$$

The wavefunction of Eqn.13 is not normalised, but of course it is generally recognised that the length of the state-vector in Hilbert space has no physical meaning. Hence normalising $\Psi_{after}$ is merely a mathematical convenience. Superposing the wavefunctions of the two histories and normalising appears to be the novel step in our calculation, although it is necessitated by the *Ansatz* that the duration of the impact has two values. In fact, it has been assumed that this step is correct since the early days of quantum mechanics. Schrödinger's cat is commonly attributed the state,

$$|\text{cat}\rangle = \frac{1}{\sqrt{2}}|\text{live cat}\rangle + \frac{1}{\sqrt{2}}|\text{dead cat}\rangle \quad (14)$$

and unquestionably a dead cat and a live cat evolve according to different Hamiltonians.

It is not clear that the coefficients $\alpha(t_0)$ and $\beta(t_0)$ in Eqn.13 are correct. There exist arguments that different histories should be added with their density matrix elements, or probabilities, as weighting coefficients. That is, the coefficients $\alpha(t_0)$ and $\beta(t_0)$ in Eqn.13 should be replaced by $\alpha(t_0)\alpha^*(t_0)$ and $\beta(t_0)\beta^*(t_0)$. However, this is not consistent with Eqn.14. Be that as it may, we have performed calculations with this replacement and the outcome is the same.

Numerical calculations show that Eqn. 13 leads to localisation after some large number of collisions. However, we do not pursue this here with the ideal gas environment, because of entanglement. If the outcome of the interaction depends on the state of the molecule, then the outcome for the incident ideal gas atom does too, and the states of the molecule and the ideal gas atom become entangled. After thirty or forty collisions, the molecule is entangled with thirty or forty gas atoms. The density matrix of Eqn.11 is $2 \times 2$; entanglement with a single gas atom requires a $4 \times 4$ density matrix, and entanglement with $n$ gas atoms requires a $2^{n+1} \times 2^{n+1}$ density matrix. Such a large density matrix is difficult to present and to read. There are ways of reducing the dimensionality so as to present a low-dimensionality density matrix in which localisation might be evident. However, in general, the entangled wavefunctions are not factorisable, so that it is not possible to reduce the dimensionality of the density matrix without losing information. For example, the decoherence programme relies on reducing the dimensionality of the density matrix by taking the trace over the environment. Tracing over the environment leads to false localisation, in Pearle's (1997) sense. Any model in which the dimensionality of the density matrix is reduced is liable to be suspected by referees of explicitly or implicitly tracing over the environment. That would yield, not true localisation through physical processes, but false localisation through mathematical approximation. In the next section, we sidestep this problem by eliminating the environment from the calculation.

*2.3 Repeated collisions of two molecules*
Consider two molecules *A* and *B* each describable by Eqns 1 to 11. They may be placed in some sort of trap, so that they collide with each repeatedly. If each is described by Eqn.4, then the pair is described by the product wavefunction,

$$\Psi_{AB} = \Psi_A \Psi_B \quad (12)$$

The density matrix in the spatial basis for the pair is

$$\rho_S = \begin{pmatrix} (\alpha_A\alpha_B)^*\alpha_A\alpha_B & (\alpha_A\alpha_B)^*\alpha_A\beta_B & (\alpha_A\alpha_B)^*\beta_A\alpha_B & (\alpha_A\alpha_B)^*\beta_A\beta_B \\ (\alpha_A\beta_B)^*\alpha_A\alpha_B & (\alpha_A\beta_B)^*\alpha_A\beta_B & (\alpha_A\beta_B)^*\beta_A\alpha_B & (\alpha_A\beta_B)^*\beta_A\beta_B \\ (\beta_A\alpha_B)^*\alpha_A\alpha_B & (\beta_A\alpha_B)^*\alpha_A\beta_B & (\beta_A\alpha_B)^*\beta_A\alpha_B & (\beta_A\alpha_B)^*\beta_A\beta_B \\ (\beta_A\beta_B)^*\alpha_A\alpha_B & (\beta_A\beta_B)^*\alpha_A\beta_B & (\beta_A\beta_B)^*\beta_A\alpha_B & (\beta_A\beta_B)^*\beta_A\beta_B \end{pmatrix} \quad (13)$$

If we let the two molecules interact as in the previous section with $t_2 = t_1$, each evolves according to Eqn.10, no localisation is observed, and the time or ensemble averaged density matrix for the pair becomes

$$\overline{\rho}_S = \begin{pmatrix} \tfrac{1}{4} & 0 & 0 & 0 \\ 0 & \tfrac{1}{4} & 0 & 0 \\ 0 & 0 & \tfrac{1}{4} & 0 \\ 0 & 0 & 0 & \tfrac{1}{4} \end{pmatrix} \quad (14)$$

However, using Eqn. 13 with $t_2 \neq t_1$, true localisation occurs. In Figure 1, we show the behaviour of the pair of molecules with $\omega_0 = 100$, $\omega_P = 10$ and $\omega_1 = 1/1000$. First we have 40 collisions with $t_2 = t_1 = \tfrac{1}{2}$, giving the random walk. The single-molecule density matrix elements $\alpha^*\alpha$ are plotted against time in Fig.1(a). At the end of this period, at $t = 4800$, the pair density matrix is

$$\rho_S = \begin{pmatrix} 0.02 & 0.07+0.03i & 0.03-0.001i & 0.10+0.04i \\ 0.07-0.03i & 0.30 & 0.10-0.05i & 0.44-0.02i \\ 0.03+0.001i & 0.10+0.05i & 0.04 & 0.14+0.06i \\ 0.10-0.04i & 0.44+0.02i & 0.14-0.06i & 0.64 \end{pmatrix} \quad (14)$$

We switch off the interaction completely for another 40 collisions, and observe the Rabi oscillations that occur when $t_2 = t_1 = 0$ (or $\omega_P = 10$), shown at Fig.1(b). The pair density matrix evolves to

$$\rho_S = \begin{pmatrix} 0.17 & 0.10-0.21i & 0.16-0.06i & 0.02-0.24i \\ 0.10+0.21i & 0.32 & 0.17+0.16i & 0.30-0.12i \\ 0.16+0.06i & 0.17-0.16i & 0.18 & 0.11-0.22i \\ 0.02+0.24i & 0.30+0.17i & 0.11+0.22i & 0.33 \end{pmatrix} \quad (15)$$

at $t = 9600$. Now we switch on the interaction with $t_2 \neq t_1$. We use $t_1 = \tfrac{1}{8}$ and $t_2 = \tfrac{3}{8}$. Localisation occurs in a few collisions, with both molecules localising into their left-hand wells. After 40 collisions, at $t = 14400$, the pair density matrix becomes

$$\rho_S = \begin{pmatrix} 0.88 & -0.10-0.20i & 0.21+0.09i & -0.005-0.06i \\ -0.10+0.20i & 0.06 & -0.05+0.04i & 0.01+0.005i \\ 0.21-0.09i & -0.05-0.04i & 0.06 & -0.01-0.01i \\ -0.005+0.06i & -0.01-0.005i & -0.01+0.01i & 0.004 \end{pmatrix} \quad (16)$$

Switching the interaction back to $t_2 = t_1 = \tfrac{1}{2}$ gives rapid delocalisation, seen at Fig.1(c), and yielding the pair density matrix

$$\rho_S = \begin{pmatrix} 0.03 & -0.08-0.08i & -0.01+0.04i & 0.12-0.06i \\ -0.08+0.08i & 0.39 & -0.06-0.12i & -0.16+0.43i \\ -0.01-0.04i & -0.06+0.12i & 0.05 & -0.11-0.11i \\ 0.12+0.06i & -0.16-0.43i & -0.11+0.11i & 0.53 \end{pmatrix} \quad (17)$$

at $t = 19200$. Finally, going back to $t_1 = ⅛$ and $t_2 = ⅜$ gives rapid localisation again, this time into the right-hand wells in both molecules, seen in Fig.1(d). After another 80 collisions, at $t = 24000$, the localised pair density matrix is

$$\rho_S = \begin{pmatrix} 0.001 & 0.003-0.0i & -0.006+0.002i & -0.02+0.01i \\ 0.003+0.0i & 0.01 & -0.03+0.005i & -0.01+0.03i \\ -0.006-0.002i & -0.03-0.005i & 0.06 & 0.23-0.03i \\ -0.02-0.01i & -0.01-0.03i & 0.23+0.03i & 0.93 \end{pmatrix} \quad (18)$$

Thus the localisation observed here is true localisation, with the diagonal elements of the entangled density matrix going to 0 or 1, and the off-diagonal elements going to zero. It is worth noting that smaller values of $\omega_1$ result in the values becoming much closer to 0 or 1, but then the Rabi oscillations (Fig.1(b)) are not seen.

*2.4 Discussion and Conclusions*

It is difficult to be precise in stating the significance of the localisation of the wavefunction observed in Figure 1 and in the density matrices of Eqn.s 14 and 14. Hund (1927) and Woolley (1976) noted that physics cannot explain chemistry, because quantum mechanics asserts that the ground state of an enantiomorphic molecule is the superposition according to Eqn.4 of its *dextro* and *laevo* forms. Yet chemists are able to distinguish the two forms, and indeed a preparation of one form or the other may be stable even for billions of years. A good example is the persistence of the handedness of the helical molecule DNA throughout billions of years of evolution. We set up our molecular dynamics model specifically to answer this challenge. The numerical values we have used are typical for a small molecule in a condensed matter environment, if we take the unit of time in Figure 1 as about a femtosecond. The localisation of the wavefunction occurs in tens of femtoseconds. Collisions with the environment account for the localisation, provided only that the duration or strength of the perturbing interaction depends on the wave-function of the system. Localisation in space occurs if the dependence is a dependence on the spatial distribution of the wave-function. This is largely in accordance with the ideas of QSD. However, in QSD, the Lindblad operator is not calculated directly from the specification of the perturbation, but is defined by the anticipated localisation. Thus the measurement of a dynamical variable with Hermitian operator **G** is represented in QSD by the Lindblad **L** = $c$**G**. Details of the perturbation are not included. The Lindblad **L** is then included in the master equation for the evolution of a given state or in the derived equation for the evolution of an ensemble density matrix. In contrast, our localisation arises directly from the nature of the perturbation without anything that might be described as "new physics" added to the Hamiltonian, to the Schrödinger equation, or to a master equation. It is also clear that the localisation cannot be due to an inadvertent tracing over the environment.

An anonymous referee pointed out that Eqn.9 is non-linear in the original wavefunction, and therefore represents a nonlinear Schrödinger evolution. This could be regarded as "new physics", but not as directly as the addition of new non-linear terms to the Schrödinger equation would be. Another referee described the model as "dodgy". Yet in the multiple-worlds interpretation of quantum mechanics, in which no collapse of the wavefunction occurs (Everett, 1957), the different worlds may evolve under different Hamiltonians. All we are suggesting here is that they are not separate worlds, but instead superpose as the real world. The superposition results in non-unitary evolution and hence localisation.

It may be questioned whether a similar analysis could be constructed for all of the classic cases of the collapse of the wavefunction in measurement, such as the Stern-Gerlach experiment, the Young's double slit experiment, the Wilson cloud chamber, and so forth. In these cases, the interaction is with the environment and the entangled density matrix becomes vary large. This question requires further work. However, a positive answer would confirm what our simple model here suggests, that the solution to the infamous measurement problem does not require the addition of new physics to the Schrödinger evolution of the system, but merely a thorough-going adherence to the Feynman adage that the key feature of quantum mechanics is the superposition of all possible histories.

**Acknowledgements** are due to Prof. Ian Percival, Mme Isabelle Herbauts, Dr Ting Yu, Dr Peter Williams, Prof. Graham Thompson, Prof. John Papaloizou and several anonymous referees for stimulating discussions. This is not to imply that they agree with the arguments presented here.

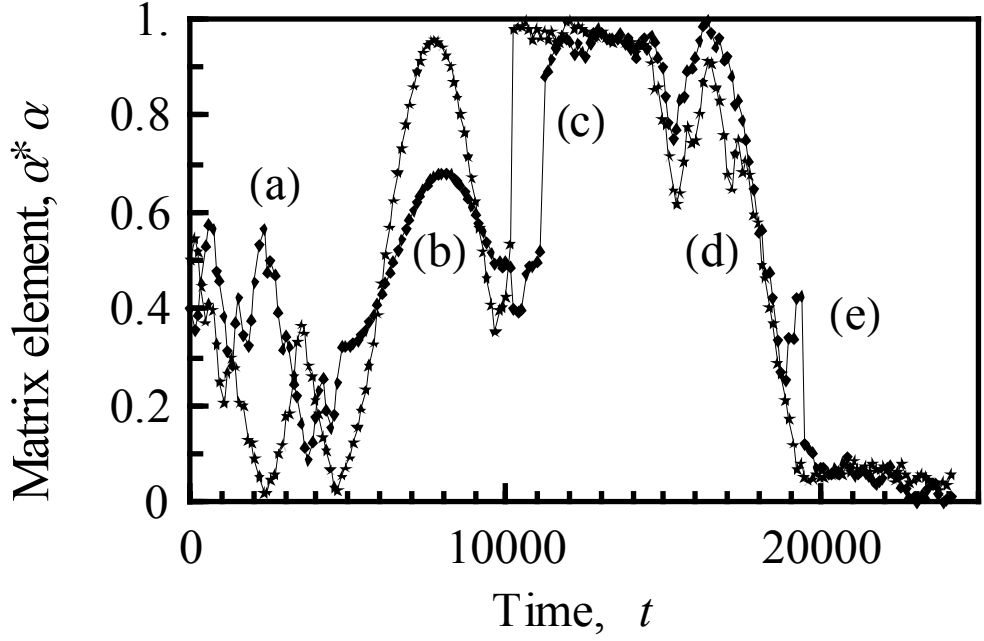

**Fig.1:** The matrix elements of the left-hand wells of each molecule, given by the squared amplitudes $|\alpha(t)|^2 = \alpha^*\alpha$, are plotted against time. The frequencies $\omega_0 = 100$, $\omega_1 = 10^{-3}$ and $\omega_P = 10$ are used. For the first 40 collisions, (a), at intervals $120 + \text{Rnd}[-20,+20]$, the impact durations are taken as equal, $t_1 = t_2 = \frac{1}{2}$. No localisation is observed and instead the matrix elements undergo a random walk. For the next 40 collisions, (b), the impact durations are set to zero so that the Rabi oscillations are observed. Then in (c) the impact durations are made unequal, with $t_1 = \frac{1}{8}$ and $t_2 = \frac{3}{8}$. The molecules localise rapidly in the left-hand wells and remain localised during 40 collisions. Then in (d) we return to $t_1 = t_2 = \frac{1}{2}$ and rapid delocalisation is seen during 40 collisions. Finally, $t_1 = \frac{1}{8}$ and $t_2 = \frac{3}{8}$ are restored for 80 collisions, (e), during which the molecules both localise again but in the right-hand wells. The pair density matrices at the end of each of these periods (a) to (e) are given in the text.